\documentclass[
aps,%
12pt,%
final,%
notitlepage,%
oneside,%
onecolumn,%
nobibnotes,%
nofootinbib,%
superscriptaddress,%
noshowpacs,%
centertags]%
{revtex4}

\begin{document}
\selectlanguage{english}

\newcounter{myfigure}

\title{\LARGE Analytical approximation of the emission line
              Fe~$K_\alpha$ in QSO's spectra}

\author{\firstname{S.V.} \surname{Repin}}
\email{repin@mx.iki.rssi.ru}
\affiliation{%
Space Research Institute of RAS \\
ul. Profsoyuznaya 84/32, 117997 Moscow, Russia
}%
\author{\firstname{V.N.} \surname{Lukash}}
\email{lukash@asc.rssi.ru}
\affiliation{%
Astrospace Center of the P.N. Lebedev Physical Institute of RAS  \\
ul. Profsoyuznaya 84/32, 117997 Moscow, Russia
}%
\author{\firstname{V.N.} \surname{Strokov}}
\email{strokov@asc.rssi.ru}
\affiliation{%
Astrospace Center of the P.N. Lebedev Physical Institute of RAS  \\
ul. Profsoyuznaya 84/32, 117997 Moscow, Russia
}%
\affiliation{%
Moscow Institute of Physics and Technology (State University) \\
Institutskiy per. 9, 141701 Dolgoprudny, Moscow region, Russia
}%


\begin{abstract}
\vskip 2cm
\begin{center}
ABSTRACT\\
\end{center}
 In spectra of many Seyfert galaxies there is a wide emission line
 of Fe $K_\alpha$. The line profile with two maxima supposes that
 the line emerges in innermost regions of an accretion disk around
a black hole, hence, it is necessary to take into account General
Relativity (GR) effects. In order to determine GR processes which
occur in active galactic nuclei (AGN) an inverse problem of
reconstructing the accreting system parameters from the line
profile has to be solved quickly. In this paper we present a
numerical approximation of the emission line Fe $K_\alpha$ with
analytical functions. The approximation is accomplished for a
range of the disk radial coordinate $r$ and the angle $\theta$
between line of sight and perpendicular to the disk and allows one
to decrease computing time by $10^6$ times in certain
astrophysical problems taking into account all GR effects. The
approximation results are available in the Internet at
\verb"http://www.iki.rssi.ru/people/repin/approx".
\end{abstract}

\maketitle


\section{Introduction}
In the past decade quite extensive X-ray observations of Seyfert
galaxies have been carried out. In a considerable body of cases
the wide emission line of Fe $K_\alpha$ ($E_0 = 6.4$~KeV) with a
peculiar two-peak profile {\cite{fabian1}-\cite{Markowitz2006}} is
observed in the active galactic nuclei (AGN) of these galaxies.
Maxima of the line are of different height, and a long red wing
may stretch up to $E\sim 3$~KeV. The Doppler line width
corresponds to matter velocities of tens of thousand kilometers
per second reaching $v \approx 10^5$~km/s for the Seyfert galaxy
MCG-6-30-15 \cite{tanaka1} and ${v = 4.8\cdot10^4}$~km/s for
MCG-5-23-16 \cite{krolik1}. Today the most reasonable explanation
seems the one which indicates that the Fe $K_\alpha$ line emerges
in inner regions of an accretion disk ($r\sim 1\div 4 r_g$) where
GR effects dominate. The Fe $K_\alpha$ line is also observed in
X-ray binaries and $\mu$QSO's. A narrow Fe~$K_\alpha$ line may
also be a part of a wide line, which cannot be distinguished from
the background.

In order to determine GR processes which occur in active galactic
nuclei (AGN) an inverse problem of reconstructing the accreting
system parameters from the line profile has to be solved quickly.
To get the line profile one has to solve numerically GR equations
of motion for photons with a wide range of initial conditions.
These computations are very time-consuming. They have already been
carried out before \cite{laor91}-\cite{zakh_rep2006}, but an
amount of effort is so unusual that it is necessary to simplify
essentially the computation procedure. Among other things, one has
to decrease the computing time and the amount of computations. In
this paper we present an approximation of the Fe $K_\alpha$ line
profile with analytical functions. The use of these functions
offers a decrease of the computing time by $10^4$ - $10^6$ times
depending on a line parameters. This procedure essentially
simplifies the solution of the inverse problem of reconstructing
AGN parameters from observational data.

In Sec. 2 we consider the procedure of calculating the Fe
$K_\alpha$ line profile within the GR framework. In Sec. 3 we
propose a method of approximating the Fe $K_\alpha$ line profile
with analytical functions. Sec. 4 deals with a numerical method
(genetic algorithm) of search for optimal approximation parameter
values. In Sec. 5 we present the approximation results and
estimate its quality. In Sec. 6 we discuss limitations of applying
the obtained approximation to the astrophysical problems. The
summary is made in Sec. 7.

\section{Calculating the line profile in GR framework}

Space-time around a rotating black hole is described with the Kerr
metrics:
$$
  ds^2 = \left( 1 - \frac{r_gr}{\rho^2} \right) dt^2 -
         \frac{\rho^2}{\Delta}\, dr^2 - \rho^2\, d\theta^2 -
$$
$$
       - \left( r^2 + a^2 + \frac{r_gra^2}{\rho^2}\sin^2\theta\right)
                   \sin^2\theta\, d\phi^2 +
         \frac{2r_gra}{\rho^2}\sin^2\theta\, d\phi dt,
$$
where $(t,r,\theta, \phi)$ are the Boyer-Lindquist coordinates.
The standard notations are introduced:
$$
     \rho^2 = r^2 + a^2\cos^2\theta, \qquad
     \Delta = r^2 - r_gr + a^2, \qquad
     r_g = 2GM,
$$
where $G$ is the gravitational constant, $a$ and $M$ are angular
momentum and mass of the black hole, respectively. Hereafter we
assume almost extreme angular momentum for the black hole
$a\approx 0.9981M$ and $G = c = 1$.

Equations of motion of free particles in the Kerr metrics are
obtained by separation of variables in the Hamilton-Jacobi
equation: \cite{Carter,wheeler,fieldtheory}
$$
   p_i p^i =
   g^{ik}\frac{\partial S}{\partial x^i}\frac{\partial S}{\partial x^k}
   = m^2,
$$
where $m$ is mass of a particle. For photons $m = 0$. The
Fe~$K_\alpha$ line profile registered by a distant observer is
obtained by solving equations of photon motion in the Kerr
metrics. After we made the set of equations dimensionless we have:
\begin{eqnarray}
   \frac{dt}{d\sigma}
                           & = &
      - a \left(a \sin^2\theta - \xi\right) +
      \frac{r^2 + a^2}{\Delta}
       \left(r^2 + a^2 - \xi a\right),
                        \label{eq1}                       \\
   \frac{dr}{d\sigma} & = & r_1,       \label{eq2}        \\
   \frac{dr_1}{d\sigma}    & = &
      2r^3 + \left(a^2 - \xi^2 - \eta\right) r +
      \left(a - \xi\right)^2 + \eta,                      \\
   \frac{d\theta}{d\sigma} & = & \theta_1,                \\
   \frac{d\theta_1}{d\sigma}
                           & = &
      \cos\theta \left(\frac{\xi^2}{\sin^3\theta} -
                       a^2 \sin\theta
                 \right),              \label{eq5}       \\
   \frac{d\phi}{d\sigma}   & = &
      - \left(a - \frac{\xi}{\sin^2\theta}\right) +
      \frac{a}{\Delta}
           \left(r^2 + a^2 - \xi a \right),
                     \label{eq6}
\end{eqnarray}
where $\sigma$ is an affine variable, $\eta$ and $\xi$ are
constants that define a trajectory of a particle. They are
expressed through quantities conserving on the trajectory: the
particle energy at infinity $E$, the projection of angular
momentum on the $z$-axis ($\theta=0$) $L_{z}$ and the Carter
\cite{Carter} separation constant\footnote{$Q=p_{\theta}^{2}+
\displaystyle\cos^{2}{\theta}\left(\frac{L_{z}^{2}}{\sin^{2}\theta}-
a^{2}M^{2}E^{2}\right)$, where $p_{\theta}$ is the
$\theta$-component of 4-momentum.} $Q$, viz. $\eta = Q/M^2E^2$ and
$\xi = L_z/ME$. In the set of equations (\ref{eq1})-(\ref{eq6})
the angular momentum $a$, the coordinates $t$ and $r$ are measured
in the units of the black hole mass. Respectively, $\Delta$
measured in the units $M^{2}$ is related to the dimensionless $r$
and $a$ as follows: $\Delta=r^2-2r+a^2$. The extra variables $r_1$
and $\theta_1$ are introduced for the set not to have
singularities, and their physical meaning is not important to us.
Two first integrals of the system
\begin{eqnarray}
  \epsilon_1 & \equiv & r_1^2 - r^4 -
      \left(a^2 - \xi^2 - \eta\right) r^2 -
      2\left[\left(a - \xi\right)^2 + \eta \right] r +
      a^2\eta = 0,                  \\
  \epsilon_2 & \equiv & \theta_1^2 - \eta - \cos^2\theta
      \left(a^2 - \frac{\xi^2}{\sin^2\theta}\right) = 0,
                    \label{eq8}
\end{eqnarray}
are used to control the calculation accuracy and avoid
accumulation of integrational errors. Namely, quantities
$\epsilon_1$ and $\epsilon_2$ have to be smaller than $10^{-8}$ in
the end of the trajectory. The method of solving equations
\mbox{(\ref{eq1})--(\ref{eq6}),} results of simulations, as well
as the derivation of the equations are available in papers
\cite{zakh_rep2,zakharov1,zakh_rep1}.

To solve the set (\ref{eq1})--(\ref{eq6}) numerically one has to
set initial conditions. We assume that particles in the disk move
along circular orbits. The disk is in an equatorial plane,
optically thick and radiates monochromatic photons with the energy
$E_{0}$ isotropically in the laboratory reference frame comoving
with the disk. We assume the disk to be opaque, i.e. the disk
absorbs a photon crossing the disk plane. In other words, a photon
emitted from one side of the disk cannot be registered from the
other side. A profile of the line radiated by a thin ring with ${r
= r_0 = const}$ is obtained by setting initial conditions at $r =
r_0$ and collecting at infinity the photons, which come to a
distant observer in the direction $\theta$. A profile of the line
radiated by the entire disk can be obtained by integrating the
intensity with respect to the radial coordinate.

Thus, given the disk and the direction $\theta$, the line profile
normalized at the unity in the maximum is the dependance of the
intensity on the photon energy at infinity. The energy is measured
in the units of the laboratory energy $E_0$:
\begin{equation}
     K_\alpha = K_\alpha\left( x \right)
     \label{KalphaDefinition}
\end{equation}
where
\begin{equation}
     x = \frac{E}{E_{0}}.
     \label{NormEnergy}
\end{equation}

\begin{figure*}[t!]
  \setcaptionmargin{5mm}
  \onelinecaptionsfalse
  \centerline{
  \includegraphics[width=7cm]{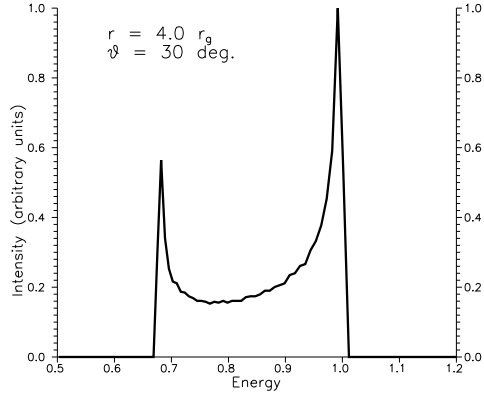}
              }
  \caption{Typical Fe~$K_\alpha$ line profile.
           Radial coordinate value \mbox{$r = 4\,r_g$}, angle
           between line of sight and perpendicular to the disk
           $\theta = 30^\circ$. Photon energy in reference frame comoving with the disk
           is taken for unity. Small-scale fluctuations are
           due to statistical reasons.}
  \label{FeKalphaTypical}
\end{figure*}

A typical Fe~$K_\alpha$ line profile registered by a distant
observer with ${\theta=30^{\circ}}$ from a ring of the radius
${r_{0}=4r_{g}}$ is shown in Fig.~\ref{FeKalphaTypical}.

\section{Line approximation}

\begin{figure*}[t!]
  \setcaptionmargin{5mm}
  \onelinecaptionsfalse
  \centerline{
  \includegraphics[width=7cm]{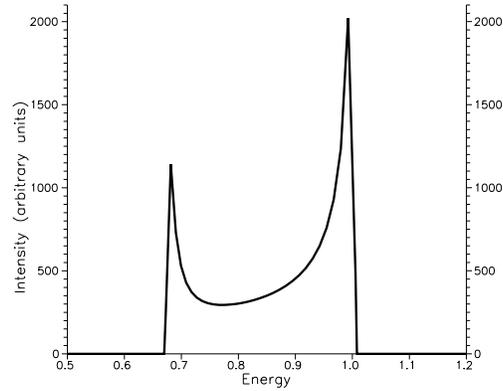}
              }
  \caption{Plot of function (\ref{eq10}) with $a_1 = 0.011181$, $\alpha_1 = 1.73588$,
           $a_2 = 0.000194$, $b_2 = 85.16571$, $\alpha_2 = 0.731736$,
           $X_1 = 0.6711022$, $X_2 = 1.005738$. Parameter values correspond to
           approximation of the curve shown in Fig.~\ref{FeKalphaTypical}.
           In this Fig. normalization is not taken into account.}
  \label{AnalyticalTypical}
\end{figure*}
If we take a look at the typical Fe $K_\alpha$ line profile (see
Fig.~\ref{FeKalphaTypical}) we notice that each maximum has both
sharp and gentle slope. Therefore, it is necessary to find an
analytical function with this property.

As it is well-known from mathematical analysis, the function $f(x)
= \exp(-a/x)$ tends to zero as $x \to 0$, moreover, it tends
faster than any power-law function does. Besides, $f(x) \to 1$ as
$x \to \infty$. The power-law function $g(x) = x^{-\alpha}$,
$\alpha
> 0$, on the contrary, tends to infinity as $x \to 0$, and to zero
as $x \to \infty$. Then their product
\begin{equation}
   y(x) = f(x)\cdot g(x) =
          \exp\left(-\frac{a}{x}\right) \cdot \frac{1}{x^\alpha}
\end{equation}
has all the required properties. On one hand, if $a$ is large
enough the latter function has a sharp slope (since the function
behavior is $\exp(-a/x)$ if $a$ is large), on the other hand the
function has a gentle power-law slope (since $\exp(-a/x) \sim 1$
if $x$ is large).

Writing the similar terms for each of the two maxima of the line
we obtain an approximation formula:
\begin{equation}
   y(x) = e^{^{-\,\frac{\displaystyle a_1}{\displaystyle\mathstrut x - X_1}}}
         \cdot \frac{\displaystyle 1}{(x - X_1)^{\alpha_1}}
         \,\,+\,\,
          e^{^{-\,\frac{\displaystyle a_2}{\displaystyle\mathstrut X_2 - x}}}
         \cdot \frac{\displaystyle b_2}{(X_2 - x)^{\alpha_2}},
         \qquad  X_1 < x < X_2,
         \label{eq10}
\end{equation}
which comprises seven parameters. The parameters $a_1$ and $a_2$
determine sharpness of the line maxima, $X_1$ and $X_2$ determine
approximate positions of both maxima, $\alpha_1$ and $\alpha_2$
determine the function behavior between the maxima, and $b_2$ sets
the relative heights of the maxima. The coefficient $b_1$ may be
set equal to unity. An example of a non-normalized plot of the
function is shown in Fig.~\ref{AnalyticalTypical}.

\section{Numerical method. Genetic algorithm.}
The approximation quality can be estimated by calculating a sum of
squared deviations of the function (\ref{eq10}) from a line
profile, which is obtained by solving the set
\mbox{(\ref{eq1})--(\ref{eq6})} and shown in
Fig.~\ref{FeKalphaTypical}. In other words, to estimate the
approximation quality it is required to find a minimum of the
function:
\begin{equation}
   F = \sum\limits_{~i\, =\, n_{_1}}^{n_{_2}}
       \left[ \frac{y\left( x_i \right)}{y_{max}} -
         K_\alpha\left( x_i \right)
       \right]^2,
   \label{approx}
\end{equation}
in the 7-dimensional parameter space ($a_1$, $X_1$, $\alpha_1$,
$a_2$, $X_2$, $b_2$, $\alpha_2$). Here $x_i$ are selected points
of the normalized energy (\ref{NormEnergy}), $y(x_i)$ is the
approximation from (\ref{eq10}), $y_{max}$ is the maximal value
from numbers $y(x_i)$, $K_\alpha(x_i)$ is the solution
(\ref{KalphaDefinition}) obtained from the set
(\ref{eq1})--(\ref{eq6}), and $n_1$ and $n_2$ are the minimal and
maximal index values when $K_\alpha(x_i) \ne 0$ (one can see in
Fig.~\ref{FeKalphaTypical} that on the right and left sides the
plot has zero segments; accurate meaning of the variables $n_1$
and $n_2$ is to be cleared below). In such cases a domain of the
7D space is usually covered with a grid and the function $F$
values are calculated in nodes of the grid. However, it is
practically impossible to accomplish the task with exhaustive
search. For minimally acceptable approximation accuracy the grid
should be fine enough: from 0 to 2 with step 0.01 in $(a_1, a_2)$,
from 0 to 2.5 with step 0.01 in $(\alpha_1, \alpha_2)$, from 0 to
200 with step 1 in $b_2$. The values $(X_1, X_2)$ are known
approximately (they are positions of the maxima), however, even
close to these values a grid with at least 100 nodes is required.
Thus, for the acceptable approximation the whole number of the
grid nodes (possible parameter combinations) is over $10^{15}$,
which is beyond computing capability.

For multidimensional problems like this one may apply the genetic
algorithm, efficiency of which increases with an increase of
number of dimensions. For the algorithm to work efficiently just
continuity and relative smoothness is required. Using the
algorithm does not guarantee that we find the exact minimum,
however, this algorithm finds this minimum with quite high
probability.

The idea of the algorithm is taken from biology: the fittest
survive, the weakest die. The domain of the multidimensional space
is first covered with a sufficiently fine grid, and nodes of the
grid are encoded with consequent binary natural numbers in each
coordinate. This binary number with a certain number of zeros and
units (bits) is called a gene. (e.g. 00101101). Each coordinate
gets its own number of bits. For instance, if the grid has 256
nodes with respect to some dimension, the genes corresponding to
this dimension contain 8 bits (one of the genes, viz. the 46th, is
mentioned above). The number of genes for each node is the same as
the number of the space dimensions. A chromosome of a certain node
is a system of the genes which are written consequently (in the
parameters' order) without spaces (e.g. if 010 and 1001 are genes
then 0101001 is a chromosome). The number of bits in a chromosome
is equal to the sum of bits of the constituent genes. Value of a
chromosome is defined as the value of the function $-F$ in a node
corresponding to this chromosome.

The essence of the algorithm is the technique with which the
chromosome with maximal value, i.e. the minimum of the function
$F$, is searched. At the initial instant some number of
chromosomes is randomly selected in the multidimensional space and
put in descending order of their values. Then the operation of
crossover is performed. To do this pairs of chromosomes are
randomly selected; the higher a chromosome value the higher the
probability to crossover. Crossover is an operation of cutting
each of the two chromosomes at a random point and exchanging the
cut parts. For the newly born chromosomes their values are
calculated. Then the entire population, parents and their
off-spring, are put in descending order again, and the weakest
chromosomes are omitted so that the number of chromosomes remains
the same. After several generations the fittest chromosome
corresponds with high probability to the minimum of the
function~$F$.

Note that just like in biology chromosomes can mutate. Mutation is
a random inversion of a bit in a chromosome. To fulfil the
procedure we make the chromosome mutate after crossover, but
before calculating its value. Namely, consequently scanning all
its bits we invert each of them with very small probability (0.001
or smaller). If a generation contains 200 chromosomes and each is
20 bits long then after mutation 4 bits (of $20 \cdot 200 = 4000$)
will be inverted. This procedure allows one to refresh the
generation and speed up the minimum search. See e.g.
\cite{Genetic1,Genetic2} for detailed description of the genetic
algorithm.

   The function (\ref{approx}) is calculated at $n$ fixed points
$x_i$ uniformly distributed in logarithmic scale. We should first
choose an interval between the points
\begin{equation}
  W = \frac{1}{n}\left( \lg E_H - \lg E_L\right) =
  \frac{1}{n} \cdot \lg \frac{\displaystyle E_H}
                             {\displaystyle E_L},
  \label{W}
\end{equation}
where $E_L$ is the minimal value of the energy, $E_H$ is the
maximal value of the energy, $n$ is number of the intervals, and
fix points of the interval separation
\begin{equation}
  \lg \xi_k = \lg E_L + kW, \qquad\qquad  k = 0,1,\dots, n.
  \label{xi}
\end{equation}
 The points $x_i$, in which we calculate values of the function $F(x)$, are
in the middle between the separation points:
\begin{equation}
  x_{k+1}\, =\, \frac{\xi_k + \xi_{k+1}}{2}\, =\,
  \frac{1 + 10^W}{2} \cdot 10^{\lg E_L + kW},
  \qquad\qquad  k = 0,1,\dots, n-1.
  \label{x_i}
\end{equation}
The values of the variables $E_L$ and $E_H$ should be chosen in
such a way that the spectral line were within the segment
$[E_L,E_H]$. Even better if the segment covers the zero segments
as well. To calculate values of the function $F$ (\ref{approx})
not all values $x_i$ are required, just the ones where
$K_\alpha(x_i) > 0$. In other words, if $n_1$ and $n_2$ stand for
the minimal and maximal values of $k$ where $K_\alpha(x_i)> 0$ we
use them in (\ref{approx}) as summation limits.

Thus, the function values are calculated only at the
aforementioned points $x_i$, and we let $F(x) = 0$ as $x <
x_{n_1}$ and $x > x_{n_2}$. To perform the calculations the
function $y(x)$ should be normalized by dividing by the maximal
value at the points $x_i$. The $F(x)$ function behavior between
the points $x_i$ is not considered. Among other things, one cannot
guarantee that the functions $F(x)$ and $K_\alpha\left( x \right)$
have the maximum at the same point. Note that the interval width
$W$ in (\ref{W}) also determines the precision of the function
value localization of $F(x)$ and $K_\alpha(x)$ on the $x$-axis. In
other words, the values of these functions are considered to be
constant between the points $\xi_k$ and $\xi_{k+1}$.

The minimum of the function (\ref{approx}) was searched at several
stages. At first, the minimum of $F(x)$ was found 10 times with
the constant parameter intervals on a quite rough grid (there were
1000~chromosomes and 80 generations). The best result of the ten
was used to center the domain in the 7-dimensional space. After
that the minimum search was performed again, but on a finer grid.
Then the process was executed two more times. The result of the
last iteration was accepted as the minimum of $F(x)$. Note that
such a cumbersome procedure is required, because the function
$F(x)$ has a number of accessory minima which are hard to
distinguish from the main minimum (this is the reason why gradient
methods are not applicable here). Using the genetic algorithm does
not guarantee that we find the "very" main maximum. However, from
our experience we can say that even if in some cases the genetics
was able to find only an accessory maximum it yielded an adequate
approximation.

\section{Simulation results}

\begin{figure*}[t!]
  \setcaptionmargin{5mm}
  \onelinecaptionsfalse %
  \centerline{
  \includegraphics[width=15cm]{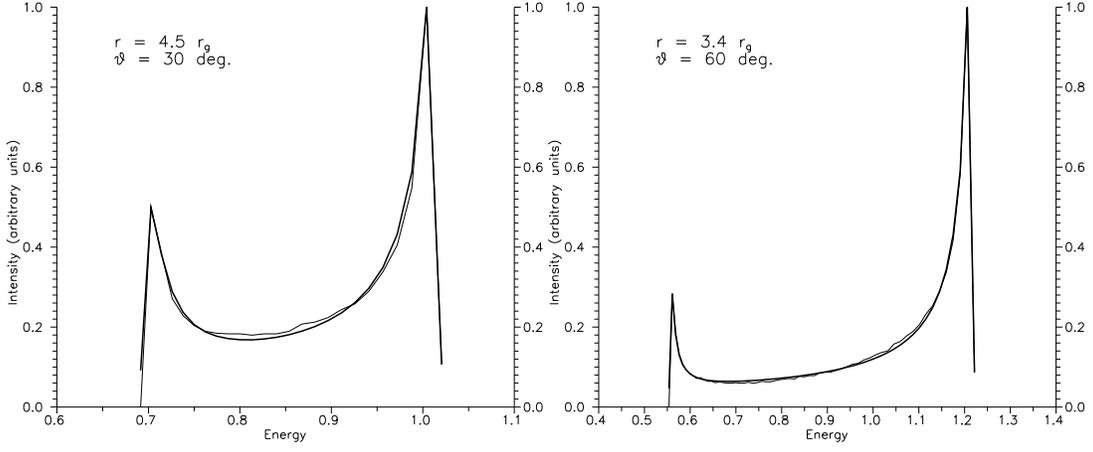}
              }
  \caption{Typical result of approximating the line Fe $K_\alpha$.
  The initial theoretical curve is plotted with a
thin line, and the result of its approximation is plotted with a
bold line. The values of the radial coordinate $r$ and an angle
$\theta$ between line of sight and perpendicular to the disk are
written on each plot. }
  \label{FeKalphaApp2}
\end{figure*}

In Fig.~\ref{FeKalphaApp2} the most typical results of the
approximation for two sets of initial parameters $r = 4.5\,r_g$,
$\theta = 30^\circ$ and $r = 3.4\,r_g$, $\theta = 60^\circ$  are
shown. The initial theoretical curve obtained by solving
numerically the set (\ref{eq1})--(\ref{eq6}) is plotted with a
thin line, and the result of its approximation is plotted with a
thick line. Theoretical curves in this and other cases are wavy,
but these fluctuations are statistical \cite{zakh_rep1} (shot
noise) and not physical. From Fig.~\ref{FeKalphaApp2} one can see
that the theoretical and approximation lines are quite close to
each other and cannot be distinguished with a naked eye.

\begin{figure*}[t!]
  \setcaptionmargin{5mm}
  \onelinecaptionsfalse %
  \centerline{
  \includegraphics[width=15cm]{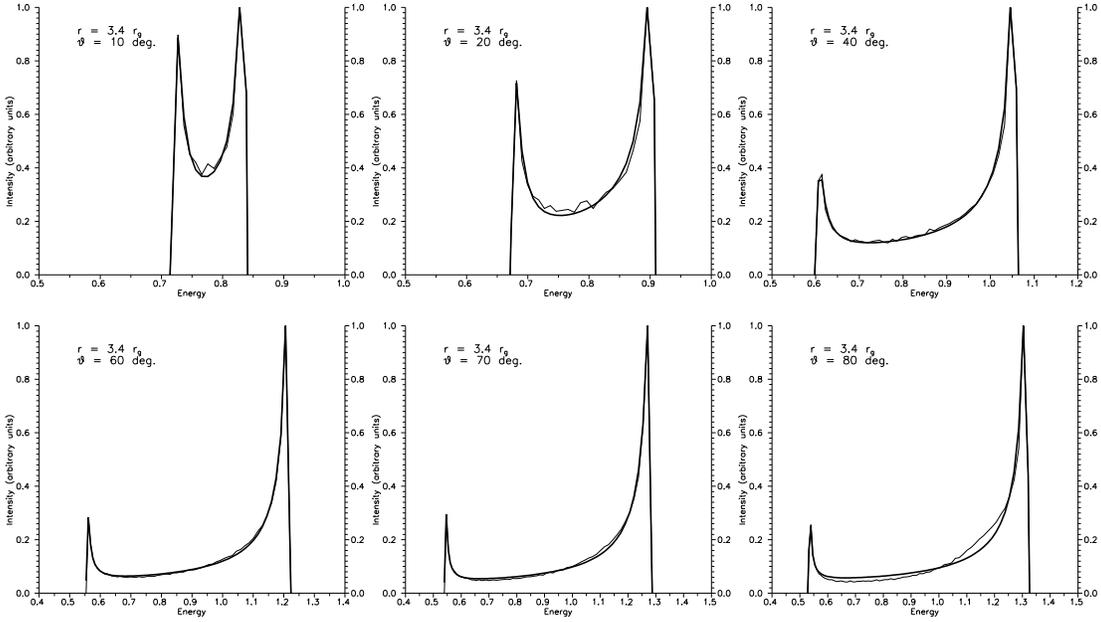}
              }
  \caption{Approximation results for the fixed value of the radial
  coordinate $r = 3.4\,r_g$ and different values of an angle
  between line of sight and perpendicular to the disk from $\theta = 10^\circ$ to
  $\theta = 80^\circ$. $x$-axis scale is different for different
  plots.}
  \label{FeKalphaApp5}
\end{figure*}

The best approximation results are impossible to be shown on the
plot at reasonable scale, because the discrepancy between the
curves is about a thin line thick. One can imagine that one of
these plots (with two curves) is shown in
Fig.~\ref{FeKalphaTypical} or~\ref{AnalyticalTypical}, but because
of the tiny discrepancy between the curves we cannot see them
separately. For $r = 4.2\,r_g$ and $\theta = 60^\circ$ the curves
deviate from each other at any point less than 0.5\% from the
maximal value.

The results of approximating the line Fe $K_\alpha$ for the fixed
value of the radial coordinate and different values of the disk
inclination angle are shown in Fig.~\ref{FeKalphaApp5}. One of the
plots with $\theta = 60^\circ$ shown in Fig.~\ref{FeKalphaApp2}
can also be added herein. As one can see from the plots the
approximation is adequate for all disk inclination angles up to
$\theta = 80^\circ$. In the latter case ($\theta = 80^\circ$) the
proposed approximation model may become inadequate, because the
lensing effects play a big role, and one or two extra maxima
\cite{zak_rep2003_aa}  may appear at some values of $r$. However,
it is possible to apply the approximations for practical problems
which do not require high accuracy. If one wants to obtain a more
accurate approximation for curves with $\theta = 80^\circ$ one can
add to the expression (\ref{eq10}) a quadratic function with a
negative leading coefficient. This extra term will describe an
extra detail in the spectrum (a detail similar to that in
Fig.~\ref{FeKalphaApp5} for $\theta = 80^\circ$ close to $E =
1.15$). For $\theta \ge 85^\circ$ the proposed model is
inadequate.

\begin{figure*}[t!]
  \setcaptionmargin{5mm}
  \onelinecaptionsfalse %
  \centerline{
  \includegraphics[width=15cm]{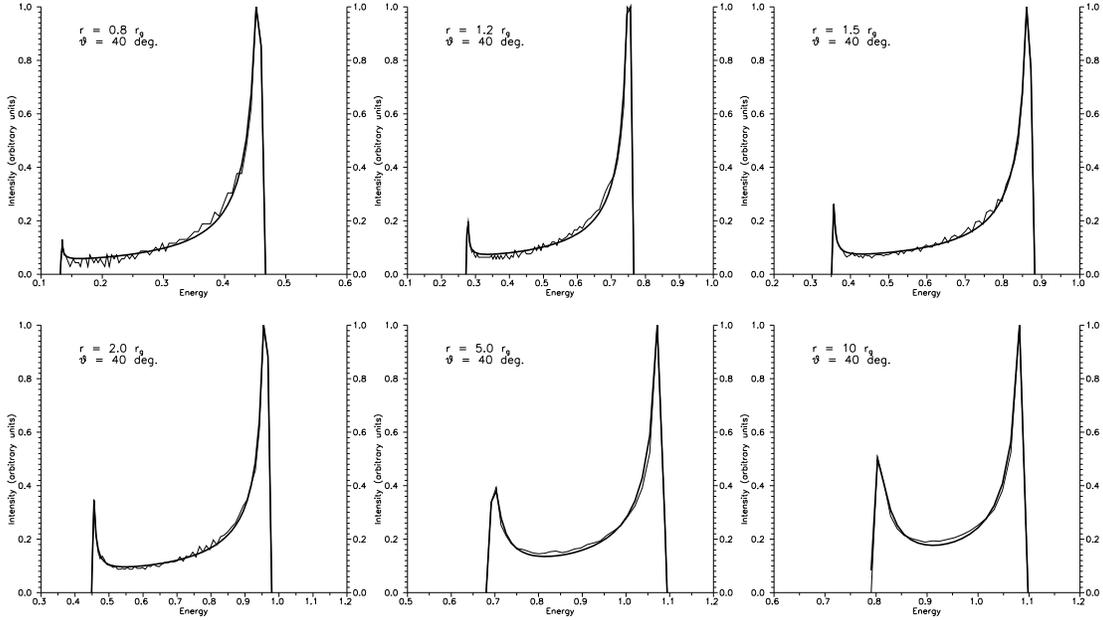}
              }
  \caption{Approximation results for the fixed value $\theta = 40^\circ$ of an angle between
  line of sight and perpendicular to the disk and
  different values of the radial coordinate from $r=0.8\,r_g$ to
  $r=10\,r_g$. $x$-axis scale is different for different
  plots.}
  \label{FeKalphaApp6}
\end{figure*}

In Fig.~\ref{FeKalphaApp6} the approximation results for the fixed
value of the disk inclination angle $\theta = 40^\circ$ and
different values of the radial coordinate $r$ are shown. As one
can see from the plots the approximation is adequate for a wide
range of the radial coordinate values. Note that the approximation
remains adequate at the very boundaries of the interval, i.e. for
$r = 0.8\,r_g$ and $r = 10\,r_g$, and some statistical
fluctuations at $r = 0.8\,r_g$ do not make it worse. Extrapolating
the results we can guess that even for $r > 10\,r_g$ the
approximation (\ref{eq10}) is quite reliable.

In Tab.~\ref{table1} values of the parameters $a_1$, $\alpha_1$,
$a_2$, $b_2$, $\alpha_2$, $X_1$, $X_2$ are given for some values
of the radial coordinate $r$ and the disk inclination angle
$\theta$. Some technical information, for which the parameter
values were obtained, are given in Tab.~\ref{table2}. The detailed
information on all the parameter values for $0.7\,r_g < r <
10\,r_g$ and $10^\circ < \theta < 80^\circ$ is available in the
Internet at \verb"http://www.iki.rssi.ru/people/repin/approx".

\begin{table}[t!]
\setcaptionmargin{5mm}
\onelinecaptionsfalse %
\begin{tabular}{|ccccccccc|}
 \hline
    &          &       &        &             &        &        &        &               \\
$r$ & $\theta$ & $a_1$ & $\alpha_1$  & $a_2$  & $b_2$  & $\alpha_2$  & $X_1$  & $X_2$    \\
    &          &       &        &             &        &        &        &               \\
 \hline
    &          &       &        &             &        &        &        &               \\
 ~1.0~ & 80 & 0.046261 & 0.15376 & 0.024139  & 97.82825  & 1.304885 & 0.1474317 & 1.416187 \\
 3.4 & 10 & 0.011971 & 1.7246  & 0.000971  & 37.33291  & 0.766634 & 0.7158835 & 0.8389752 \\
 3.4 & 20 & 0.010876 & 1.72776 & 0.000365  & 84.92977  & 0.672685 & 0.6712679 & 0.9066637 \\
 3.4 & 30 & 0.008042 & 1.41    & 0.00064   & 35.70628  & 0.701063 & 0.6354619 & 0.9801457 \\
 3.4 & 40 & 0.011937 & 1.569   & 0.000698  & 62.20945  & 0.730865 & 0.6015416 & 1.059468  \\
 3.4 & 50 & 0.011593 & 1.7134  & 0.001139  & 37.30477  & 0.768584 & 0.7159498 & 0.8390415 \\
 3.4 & 60 & 0.006658 & 1.4788  & 0.001595  & 82.66419  & 0.845563 & 0.5537415 & 1.222009  \\
 3.4 & 70 & 0.003384 & 1.4204  & 0.005798  & 84.84075  & 0.939793 & 0.5395592 & 1.287379  \\
 3.4 & 80 & 0.000955 & 1.15504 & 0.004171  & 43.38763  & 0.90118  & 0.532406  & 1.321832  \\
 3.5 & 10 & 0.008442 & 1.67216 & 0.000816  & 56.68913  & 0.731884 & 0.7252888 & 0.8497934 \\
 3.5 & 20 & 0.005904 & 1.40092 & 0.001682  & 29.44844  & 0.73757  & 0.680184  & 0.9187329 \\
 3.5 & 30 & 0.012463 & 1.65552 & 0.002242  & 65.77657  & 0.756707 & 0.6386765 & 0.9931731 \\
 3.5 & 40 & 0.014688 & 1.66824 & 0.011318  & 51.58594  & 0.959569 & 0.6063802 & 1.073101  \\
 3.5 & 50 & 0.002631 & 1.28648 & 0.00097   & 40.9329   & 0.802594 & 0.5832343 & 1.145212  \\
 3.5 & 60 & 0.007254 & 1.44316 & 0.002104  & 63.847    & 0.853285 & 0.5609719 & 1.222076  \\
 3.5 & 70 & 0.006386 & 1.5494  & 0.013498  & 97.622    & 1.032985 & 0.5466044 & 1.286815  \\
 3.5 & 80 & 0.001006 & 1.33604 & 0.00075   & 77.44388  & 0.890169 & 0.5394267 & 1.320739  \\
 4.1 & 10 & 0.00591  & 1.517   & 0.000592  & 49.33127  & 0.643282 & 0.7605556 & 0.8719551 \\
 4.1 & 20 & 0.004271 & 1.29656 & 0.00042   & 31.59231  & 0.61765  & 0.7180462 & 0.9353039 \\
 4.1 & 30 & 0.006604 & 1.37288 & 0.000688  & 36.44845  & 0.648607 & 0.6786068 & 1.00382   \\
 4.1 & 40 & 0.004174 & 1.37864 & 0.001503  & 62.75788  & 0.692406 & 0.6498227 & 1.075424  \\
 4.1 & 50 & 0.001696 & 1.17712 & 0.000427  & 39.10009  & 0.685844 & 0.6245808 & 1.1435    \\
 4.1 & 60 & 0.012402 & 1.63436 & 0.002646  & 89.25638  & 0.771792 & 0.5985646 & 1.20719   \\
 6.5 & 60 & 0.019516 & 1.66296 & 0.0003    & 34.00781  & 0.859408 & 0.6902458 & 1.191818  \\
    &   &   &   &   &   &   &   &   \\
 \hline
\end{tabular}
 \caption{Approximation results for some values of radial
 coordinate $r$ and angle $\theta$ between line of sight and perpendicular
 to the disk.}
 \label{table1}
\end{table}

\begin{table}[t!]
\setcaptionmargin{5mm}
\onelinecaptionsfalse %
\centerline{
\begin{tabular}{|cccccc|}
 \hline
    &          &        &        &      &   \\
$r$ & $\theta$ & $E_L$  & $E_H$  & $\mathstrut n$  & Sum  (\ref{approx}) \\
    &          &        &        &      &   \\
 \hline
    &          &        &        &      &   \\
 1.0 & 80 & 0.1 & 1.5 & 150 & 0.0252970  \\
 3.4 & 10 & 0.2 & 1.4 & 150 & 0.0067702  \\
 3.4 & 20 & 0.2 & 1.4 & 150 & 0.0143621  \\
 3.4 & 30 & 0.2 & 1.4 & 150 & 0.0064464  \\
 3.4 & 40 & 0.2 & 1.4 & 150 & 0.0070760  \\
 3.4 & 50 & 0.2 & 1.4 & 150 & 0.0068298  \\
 3.4 & 60 & 0.2 & 1.4 & 150 & 0.0017467  \\
 3.4 & 70 & 0.2 & 1.4 & 150 & 0.0036399  \\
 3.4 & 80 & 0.2 & 1.4 & 150 & 0.0281192  \\
 3.5 & 10 & 0.2 & 1.4 & 150 & 0.0023694  \\
 3.5 & 20 & 0.2 & 1.4 & 150 & 0.0047536  \\
 3.5 & 30 & 0.2 & 1.4 & 150 & 0.0065227  \\
 3.5 & 40 & 0.2 & 1.4 & 150 & 0.0250608  \\
 3.5 & 50 & 0.2 & 1.4 & 150 & 0.0030569  \\
 3.5 & 60 & 0.2 & 1.4 & 150 & 0.0020274  \\
 3.5 & 70 & 0.2 & 1.4 & 150 & 0.0065676  \\
 3.5 & 80 & 0.2 & 1.4 & 150 & 0.0209602  \\
 4.1 & 10 & 0.6 & 1.0 &  80 & 0.0080929  \\
 4.1 & 20 & 0.6 & 1.0 &  80 & 0.0119948  \\
 4.1 & 30 & 0.5 & 1.4 & 150 & 0.0078276  \\
 4.1 & 40 & 0.5 & 1.4 & 150 & 0.0088596  \\
 4.1 & 50 & 0.5 & 1.4 & 150 & 0.0037481  \\
 4.1 & 60 & 0.2 & 1.4 & 150 & 0.0053153  \\
 6.5 & 60 & 0.3 & 1.4 & 110 & 0.0066761  \\
    &   &   &   &   &   \\
 \hline
\end{tabular}
}
 \caption{Technical information for the same values of radial
 coordinate $r$ and angle $\theta$ between line of sight and perpendicular
 to the disk as in Tab.\ref{table1}.}
 \label{table2}
\end{table}

\begin{figure*}[t!]
\setcaptionmargin{5mm}
\onelinecaptionsfalse %
    \includegraphics[width=\textwidth]{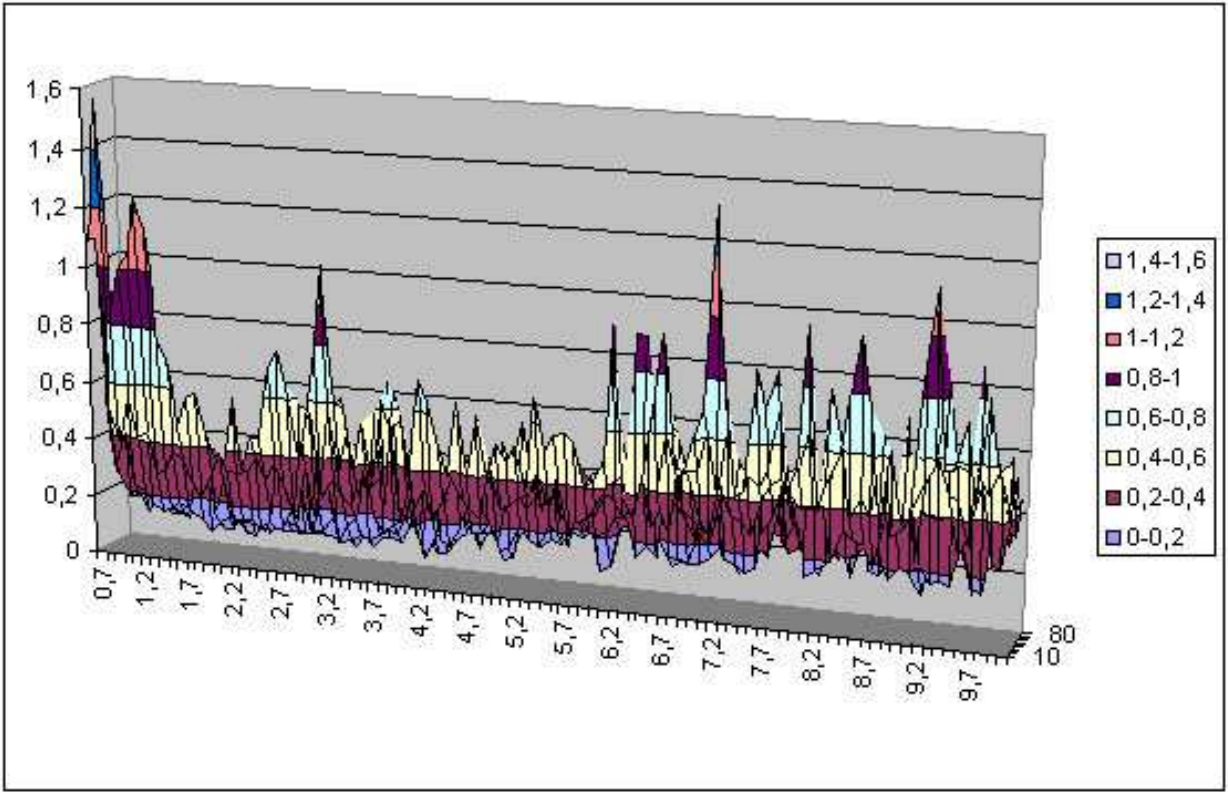}
  \caption{Approximation quality. Root-mean-square $z_1(r,\theta)$
  in significant part of the curve measured in percentage. $x$- and
  $y$-axes are variables~$r$~and~$\theta$, respectively.}
  \label{approx_quality2}
\end{figure*}

To estimate the approximation quality beside the function
(\ref{approx}) we also used
\begin{equation}
   z_1 \left(r,\theta\right) =
    \sqrt{
    \frac{\displaystyle
           \sum\limits_{i=n_{_1}}^{n_{_2}}
             \left[\mathstrut
                   \frac{y\left( x_i \right)}{y_{max}}
                    - K_\alpha\left( x_i \right)
             \right]^2
            }
         {\displaystyle\mathstrut n_2 - n_1 + 1}}\quad,
   \label{quality2}
\end{equation}
where $n_1$ and $n_2$ are the numbers of the first and last
intervals in a significant part of the curve, respectively. This
value is a root-mean-squared per interval in a significant part of
the curve. The term "significant" means that we consider only the
intervals with at least one photon. For instance, for $r =
3.4\,r_g$ and $\theta = 20^\circ$ the calculations were carried
out for $E_L = 0.2$, $E_H = 1.4$, $n = 150$. The line itself
occupies the interval $0.66 < x_i < 0.92$. The statement made
above means that while calculating the root-mean-square the
intervals with $x_i < 0.66$ and $x_i > 0.92$ were omitted.

It is convenient to measure the function
$z_1\left(r,\theta\right)$ in percentage. The plot of this
function is shown in Fig.~\ref{approx_quality2}. The values
0.2~-~0.5\% are typical. The best approximation results are for
$1.5\,r_g < r < 6\,r_g$. For higher and lower values of $r$ the
approximation is some worse, but even though the deviations higher
than 0.7\% are quite rare. For stand-alone points with $r < r_g$
and $r > 6.5\,r_g$ a more accurate approximation will be found
some day.

\begin{figure*}[t!]
  \setcaptionmargin{5mm}
  \onelinecaptionsfalse %
  \includegraphics[width=\textwidth]{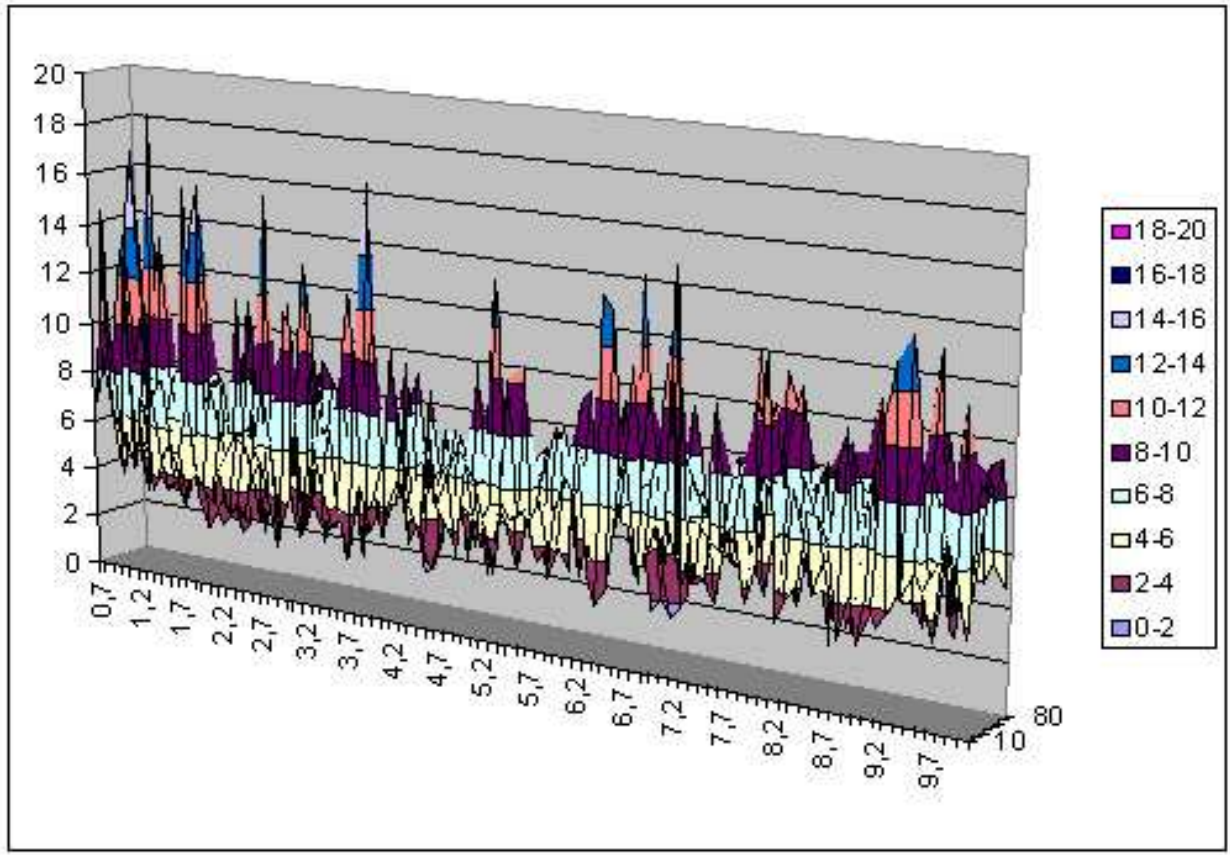}
  \caption{Approximation quality. Plot of the maximal
  discrepancy (in percentage) between curves is shown. $x$- and
  $y$-axes are variables~$r$~and~$\theta$, respectively.}
  \label{approx_quality1}
\end{figure*}

To estimate the approximation quality one can use an extra
criterion which describes the maximal discrepancy between
theoretical and approximation curves
\begin{equation}
   z_2 \left(r,\theta\right)\, = \,
   \max \left| \frac{y\left( x_i \right)}{y_{max}} -
   K_\alpha\left( x_i \right) \right|, \qquad\quad
   i = 1, \dots, n.
   \label{quality1}
\end{equation}
and measured in percentage of the maximal value of a curve. The
plot of the value is shown in Fig.~\ref{approx_quality1}. As one
can see from the plot, the maximal discrepancy is mainly 6-8\%,
and discrepancies higher than 10\% are quite rare. Besides, this
criterion has no explicit dependance on variables $r$ and
$\theta$. Intervals with maximal discrepancies are most often
located close to the blue (higher) maximum from its left side
where the curve drops sharply. Usually it is one or two points.
Thus, this criterion characterizes "the worst" point of the curve.

\section{Discussion}

As it has been noticed above, the function $F(x)$ values can be
taken only at the points $x_i$ defined by the relations (\ref{W})
and (\ref{x_i}), because between these points the function $F(x)$
values are not defined. Since it also applies to the points $x_i$
close to the maxima the used approach does not guarantee that the
function $F(x)$ reaches its maximum at the same point as
$K_\alpha\left( x \right)$. When using the approximation
(\ref{eq10}) (e.g. to model a profile of the line radiated by the
entire disk \cite{novikov,shakura,shasun}) one should first
calculate intensities at the points $x_i$ for several thin rings
and then sum up the results with a decrease of number of intervals
in the resulting curve (hence, make it rougher). This procedure
allows one to obtain a quite smooth curve which can be used to
interpret observational data.

\begin{figure*}[t!]
  \setcaptionmargin{5mm}
  \onelinecaptionsfalse %
  \includegraphics[width=\textwidth]{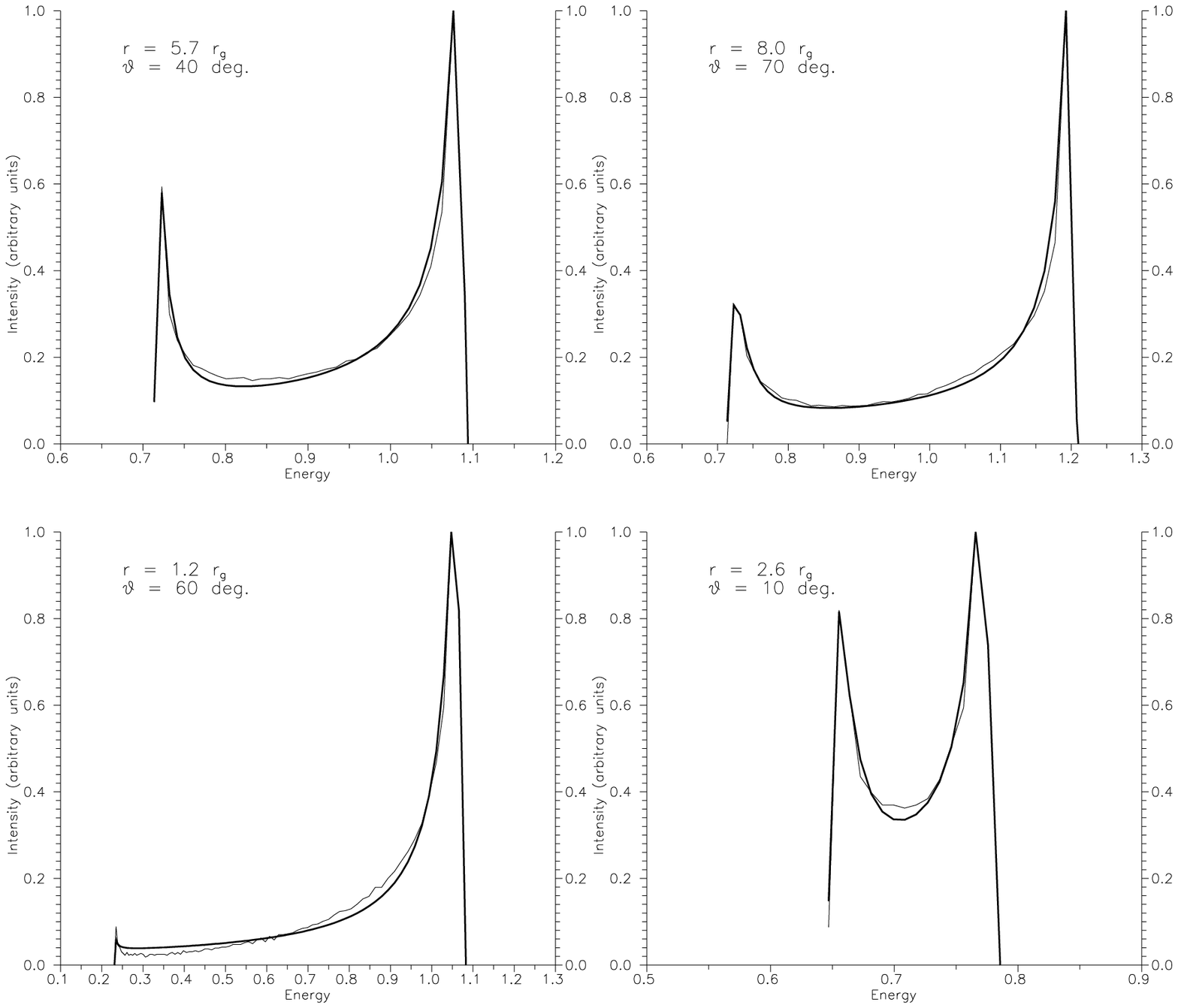}
  \caption{Examples of approximations with the highest deviation
  from theoretical curves. Values of radial coordinate $r$ and
  angle $\theta$ between line of sight and perpendicular to the disk
  are written on each plot.}
  \label{FeKalphaApp3}
\end{figure*}

In Fig.~\ref{FeKalphaApp3} examples of the approximations with the
highest deviations from the theoretical curves are shown. Although
these cases are rare, we find it important to explain reasons
which caused them. One of the reasons is improper photon energy
distribution in higher (blue) and lower (red) maxima. As a result,
a maximum consists of two values with almost the same height which
is not suitable for the approximation (\ref{eq10}). This situation
occurs for the second (red maximum) and third (blue maximum) plots
in Fig.~\ref{FeKalphaApp3}. This problem can be solved by changing
the location of the points $x_i$ close to the maxima which, in
turn, can be achieved by changing values of $E_L$, $E_H$ and $n$.
In this way one can obtain approximations of higher quality for
the curves shown in Fig.~\ref{FeKalphaApp3}, but this is quite
time-consuming. The other reason appears at the small values $r <
1.2\,r_g$. Here in a quite wide range of the disk inclination
angle $\theta$, as a matter of fact, a curve has just one distinct
maximum and the approximation (\ref{eq10}) does not work.

\section{Conclusions}

We obtained an analytical approximation of the iron emission line
profile in AGN, which allows one to determine physical parameters
of the matter of the accreting disk around a black hole by the
line profile.

According to the criterion (\ref{quality2}) in the intervals
$1.3\,r_g < r < 2.5\,r_g$ with respect to the radial coordinate
and $10^\circ < \theta \le 80^\circ$ with respect to the angle
$\theta$ as well as $3.5\,r_g < r < 6\,r_g$, $20^\circ < \theta
\le 80^\circ$ the approximation (\ref{eq10}) yields the accuracy
of 0.8\% excluding several points with higher deviation, but still
lower than 1.4\%. In the interval $0.7\,r_g < r \le 1.3\,r_g$ for
all values of $\theta$ one has the accuracy of 1.2\% excluding two
points with 1.6\%.

According to the criterion of the maximal discrepancy
(\ref{quality1}) the approximation (\ref{eq10}) yields the
accuracy from 2\% to 12\% in the entire investigated interval
$0.7\,r_g < r < 10\,r_g$. No explicit dependance of the
approximation accuracy on $r$ and $\theta$ was noticed. As an
exception there are less than ten points (mainly at $r < 4\,r_g$)
in which the discrepancy amounts to 18\%.

The application of the approximation (\ref{eq10}) in practical
astrophysical problems results in a decrease of computing time by
$10^4 - 10^6$ times.

One of us (SVR) expresses his gratitude to Prof.~E.V.~Starostenko,
Dr.~R.E.~Beresneva and Dr.~O.N.~Sumenkova for the possibility to
work intensively and fruitfully on the considered problem and to
Prof.~A.F.~Zakharov for fruitful discussions. VNL and VNS are
grateful to Russian Fond of Basic Research for partial support of
this paper, Grants 04-02-17444, 07-02-00886. VNS thanks UNK FIAN
for support.

\end{document}